\documentclass{iopart}

\usepackage{graphicx}
\usepackage{subfig}

\usepackage{amsopn}
\usepackage{epsfig}
\usepackage{breqn}
\usepackage{epstopdf}
\usepackage{bm}
\usepackage{amssymb}
\usepackage{color}
\usepackage{cite}
\usepackage{enumitem}

\usepackage{makecell}
\usepackage{geometry}
\usepackage[abbr]{harvard}
\citationstyle{agsm}
\renewcommand{\harvardand}{and}

\newcommand{\bb}{\bm}
\usepackage{algorithm,algorithmic}
\newcommand{\txtc}{\color{black}}

\begin{document}
\title{Attenuation Correction for Brain PET Imaging Using Deep Neural Network Based on Dixon and ZTE MR Images}
\author{Kuang Gong$^{1}$,  Jaewon Yang$^{2}$,  Kyungsang Kim$^1$, Georges El Fakhri$^1$,Youngho Seo$^2$, and Quanzheng Li$^1$*}
\address{$^1$Gordon Center for Medical Imaging, Department of Radiology, Massachusetts General Hospital, Boston, MA, 02114 USA}%
\address{$^2$Physics Research Laboratory, Department of Radiology and Biomedical Imaging, University of California, San Francisco, CA 94143 USA}%
\ead{li.quanzheng@mgh.harvard.edu}

\begin{abstract}
Positron Emission Tomography (PET) is a functional imaging modality widely used in neuroscience studies. To obtain meaningful quantitative results from PET images, attenuation correction is necessary during image reconstruction. For PET/MR hybrid systems, PET attenuation is challenging as Magnetic Resonance (MR) images do not reflect attenuation coefficients directly. To address this issue, we present deep neural network methods to derive the continuous attenuation coefficients for brain PET imaging from MR images. With only Dixon MR images as the network input, the existing U-net structure was adopted and analysis using forty patient data sets shows it is superior than other Dixon based methods. When both Dixon and zero echo time (ZTE) images are available, we have proposed a modified U-net structure, named GroupU-net, to efficiently make use of both Dixon and ZTE information through group convolution modules when the network goes deeper. Quantitative analysis based on fourteen real patient data sets demonstrates that both network approaches can perform better than the standard methods, and the proposed network structure can further reduce the PET quantification error compared to the U-net structure.

\end{abstract}
\section{Introduction}
Positron Emission Tomography (PET) can produce three dimensional images of biochemical processes in the human body by using specific radioactive tracers. It has wide applications in neuroscience studies, such as measurement of metabolism for brain tumor imaging, dopamine neurotransmitter imaging related to addiction, $\beta$-amyloid and tau imaging in Alzheimer's disease, translocator protein (TSPO) imaging related to microglial activation and so on. Due to various physical degradation factors, correction items, such as randoms, scatters, normalization and attenuation correction, should be included in the reconstruction process to obtain meaningful quantitative results. For attenuation correction, information from computed tomography (CT) has been treated as a reference standard to reflect the attenuation coefficients in 511 Kev after a bilinear scaling \cite{kinahan2003x}. 

Recently, PET/MR systems begin to be adopted in clinics due to MR's excellent soft tissue contrast and the ability to perform functional imaging. In addition, simultaneously acquired MR images can provide useful information for PET motion compensation \cite{catana2011mri} and partial volume correction \cite{gong2017direct}. One concern is that the MR signal is not directly reflective of attenuation coefficients, and hard to be used for attenuation correction without approximation. Many methods have been proposed to generate the attenuation map based on T1-weighted, Dixon, ultra-short echo time (UTE) or zero echo time (ZTE) MR images, which can majorly be summarized into four categories. The first category is segmentation based methods. The MR image is segmented into different tissue classes with the corresponding attenuation coefficients assigned to produce the attenuation map  \cite{martinez2009tissue,keereman2010mri,berker2012mri,ladefoged2015region,sekine2016clinical,leynes2017hybrid,khalife2017subject,yang2017evaluation}.  Another category relies on the atlas generated from prior patients' CT and MR pairs. Pseudo CT will be created by non-rigidly registering the atlas to patient MR images \cite{wollenweber2013evaluation,burgos2014attenuation,izquierdo2014spm8,yang2017quantitative}. With the availability of time-of-flight (TOF) information, emission based methods have been developed to estimate the activity image and the attenuation map simultaneously without the use of MR information \cite{defrise2012time,rezaei2012simultaneous,li2017joint}, or aided by MR information \cite{mehranian2015joint,kim2016penalized,mehranian2017mr}. Finally, there are efforts adopting machine learning based approaches to pseudo CT generation driven by prior MR and CT pairs, such as the random forest \cite{huynh2016estimating} and neural network methods\cite{han2017mr,nie2017medical,liu2017deep,leynes2017direct}. 

Over the past several years, deep neural networks have been widely and successfully applied to computer vision tasks because of the availability of large data sets, advances in optimization algorithms and emerging of effective network structures. Recently, it has been applied to medical imaging, such as image denoising \cite{wang2016accelerating,kang2016deep,chen2017low}, image reconstruction \cite{wu2017iterative,gong2017iterative} and end-to-end lesion detection \cite{wu2017end}. Several pioneering works have shown that neural networks can be employed to generate the pseudo CT images from T1-weighted MR images for the brain region, with evaluations on the pseudo CT image quality only \cite{han2017mr,nie2017medical}. Take one step further, \citeasnoun{liu2017deep} used convolutional
auto-encoder (CAE) to generate the CT tissue labels (air, bone,
and soft tissue) from T1-weighted MR images and evaluated its performance for PET images. In that work additional CT segmentation is needed and the attenuation coefficients were assigned based on tissue labels, which are not continuous. Recently \citeasnoun{leynes2017direct} combined ZTE and Dixon images to generate the pseudo CT for the pelvis region using the {\txtc{U-net structure \cite{ronneberger2015u}}}. 

In this work, we focus on using neural network based methods to predict the continuous attenuation map specifically for brain PET imaging under two scenarios:
\begin{enumerate}
\item When there are only Dixon MR images available, we adopted the U-net structure \cite{ronneberger2015u} to generate the pseudo CT images. Forty patients' data sets were used in the experiment and cross-validated to evaluate the performance. The segmentation and atlas methods based on Dixon MR images provided by the vendor were used as comparison methods;
\item  When both Dixon and ZTE MR images are available, we proposed {\txtc{a new network structure based on group convolution modules}} to more efficiently combine ZTE and Dixon information. Fourteen patient data sets with both Dixon and ZTE images were employed in the experiments. The ZTE segmentation method provided by the vendor was adopted as the comparison methods.
\end{enumerate}

The main contributions of this paper include (1) using deep neural networks to generate continuous attenuation maps for brain PET imaging;(2) proposing a new network structure to generate the attenuation maps utilizing multiple MR inputs; (3) a comprehensive quantitative comparison with the standard methods. 

\section{Method}
\subsection{PET attenuation model}
For PET image reconstruction, the measured sinogram data $\bb{y} \in \mathbb{R}^{M \times 1} $ can be modeled as a collection of independent Poisson random variables and its mean  $\bar{\bb{y}} \in \mathbb{R}^{M \times 1} $ is related to the unknown image $\bb{x} \in \mathbb{R}^{N \times 1}$ through an affine transform
\begin{equation}
\bar{\bb{y}} = \bb{P}\bb{x} + \bb{s} + \bb{r},
\label{eqn_mean}
\end{equation}
where $\bb{P} \in \mathbb{R}^{M \times N}$ is the detection probability matrix, $\bb{s} \in \mathbb{R}^{M \times 1}$ is the expectation of scattered events, and $\bb{r} \in \mathbb{R}^{M \times 1}$ denotes the expectation of random coincidences. $M$ is the number of lines of response (LOR) and $N$ is the number of pixels in image space. 
The reconstructed image quality strongly depends on the accuracy of the detection probability matrix $\boldsymbol{P}$, which can be decomposed to \cite{qi1998high}
\begin{equation}
\label{eqn_sys}
\boldsymbol{P} = \boldsymbol{N}\boldsymbol{A}\boldsymbol{B}\boldsymbol{G}\boldsymbol{R},
\end{equation}
where $\boldsymbol{G}\in \mathbb{R}^{M \times N}$ is the geometric projection matrix whose element $g_{i,j}$ denotes the probability of a photon pair produced in voxel $j$ reaching the front faces of detector pair $i$, $\boldsymbol{R}\in \mathbb{R}^{N \times N}$ models the image domain blurring effects, $\boldsymbol{B}\in \mathbb{R}^{M \times M}$ is the sinogram domain blurring matrix \cite{gong2017sinogram},  diagonal matrix $\boldsymbol{N} \in \mathbb{R}^{M \times M}$ contains the normalization effects, and diagonal matrix $\boldsymbol{A} \in \mathbb{R}^{M \times M}$ models the attenuation factors. The $i$th diagonal element of attenuation matrix $\boldsymbol{A}$ is calculated as
\begin{equation}
a_{ii} = e^{-\sum_j l_{ij}\mu_j},
\end{equation}
where $\boldsymbol{\mu}  \in \mathbb{R}^{N_{\mu} \times 1} $ is the attenuation map, $l_{ij}$ denotes the interaction length of LOR $i$ with voxel $j$. In PET/CT, CT images are used for the attenuation map generation by the bilinear scaling method \cite{carney2006method}
\begin{equation}
{\mu_j} = \left\{  \begin{array}{r@{\quad}cr} 
 9.6\e^{-5}(\mbox{HU}_j + 1000)& \mathrm{if} & \mbox{HU}_j < \mbox{Threshold}, \\  
a(\mbox{HU}_j + b) &  \mathrm{if} & \mbox{HU}_j > \mbox{Threshold}.   
\end{array}\right.
\end{equation}
Here $\mbox{HU}_j$ represents the HU units in CT voxel $j$. $a$, $b$ and $ \mbox{Threshold}$ are values depending on the energy of the CT and are given in \citeasnoun{carney2006method}.
\subsection{Pseudo CT generation using deep neural network}
{\txtc{The basic module of a convolutional neural network includes a convolution layer and an activation layer. The input and output relationship of the $i^{\small{\mbox{th}}}$ module can be denoted as
\begin{equation}
\bb{y}_i = f_i(\bb{y}_{i-1}) = g(\bb{w}_i \circledast \bb{y}_{i-1} + \bb{b}_i),
\label{convo_network}
\end{equation}
where  $\bb{y}_{i-1} \in \mathbb{R}^{N \times N\times C}$ is the module input with spatial size $N \times N$ and channel size $C$, $\bb{y}_i  \in \mathbb{R}^{N \times N\times H} $ denotes the module output with spatial size $N \times N$ and $H$ channels, $\bb{w}_i \in \mathbb{R}^{M \times M \times C \times H}$ is the convolutional filter with kernel width $M$, $\bb{b} \in \mathbb{R}^{1 \times H}$ is the bias term, $\circledast$ stands for the convolution operation, and $g$ represents the non-linear activation function. The rectified linear unit (ReLU) activation function, defined as $g(\bb{x}) = \mbox{max}(\bb{x},\bb{0})$, is employed as the activation function. To stabilize and accelerate the deep network training, batch normalization \cite{ioffe2015batch} is often added after the convolution operation. After stacking $L$ units together, the network output can be calculated as
\begin{equation}
\bb{y}_{\small{\mbox{out}}}(\bb{x}_{\small{\mbox{input}}})= f_L(f_{L-1}(...f_1(\bb{x}_{\small{\mbox{input}}}))).
\label{convo_network}
\end{equation}}}
{\txtc{In this work,}} MR images are treated as the network input and pseudo CT images are output of the network. The network is trained based on prior acquired MR and CT pairs from different patients, with the objective function
\begin{equation}
L = |\mbox{CT}_{\small{\mbox{true}}} - \bb{y}_{\small{\mbox{out}}}(\mbox{MR})|,
\end{equation}
which is the L1 norm of the difference between the ground truth CT image $\mbox{CT}_{\small{\mbox{true}}}$ and the output from the neural network $\bb{y}_{\small{\mbox{out}}}(\mbox{MR})$. We have also tried L2 norm and found that L1 norm can produce less blurred structures.

\begin{figure}[t]
\centering
\subfloat{\includegraphics[trim=0cm 3.5cm 0.5cm 2cm, clip, width=5.5in]{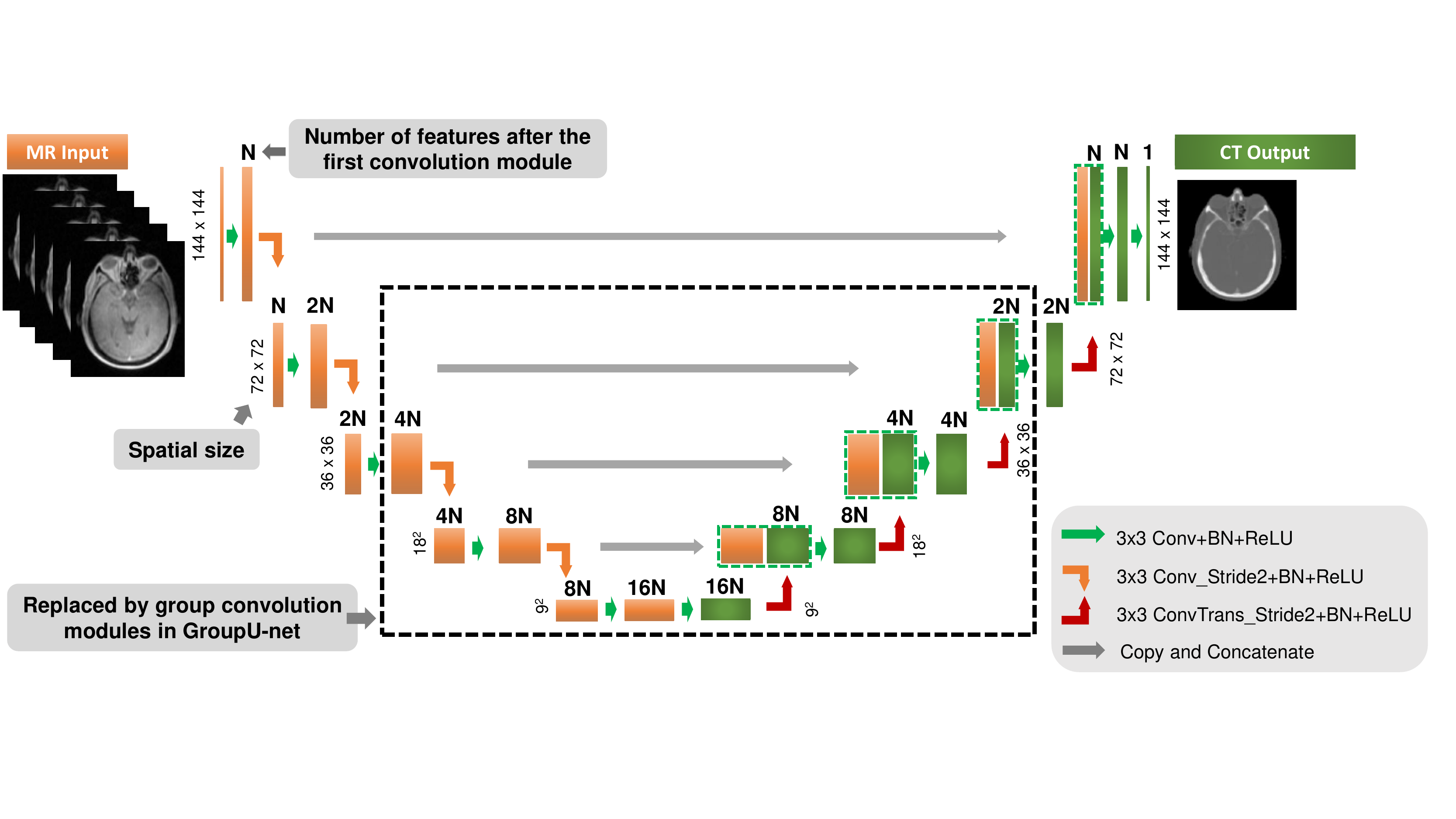}}
\caption{\small{The schematic diagram of the U-net architecture. {\txtc{Numbers on top of the module stand for the number of features in the channel. Numbers on the left size of the module indicate the spatial input size. $N$ is the number of features after the first convolution module. For the proposed GroupU-net structure, the convolution module inside the dashed box will be replaced by the group convolution module as indicated in Fig.~\ref{fig:group-conv-module}. The group module will only be used when the block input has features  $\ge 4N$. The number of groups in the group convolution module is set to be $N$.}}}}
\label{fig:u-net-structure}
\end{figure}
\begin{figure}[t]
\centering
\subfloat{\includegraphics[trim=12cm 5cm 0.5cm 2cm, clip, width=5.5in]{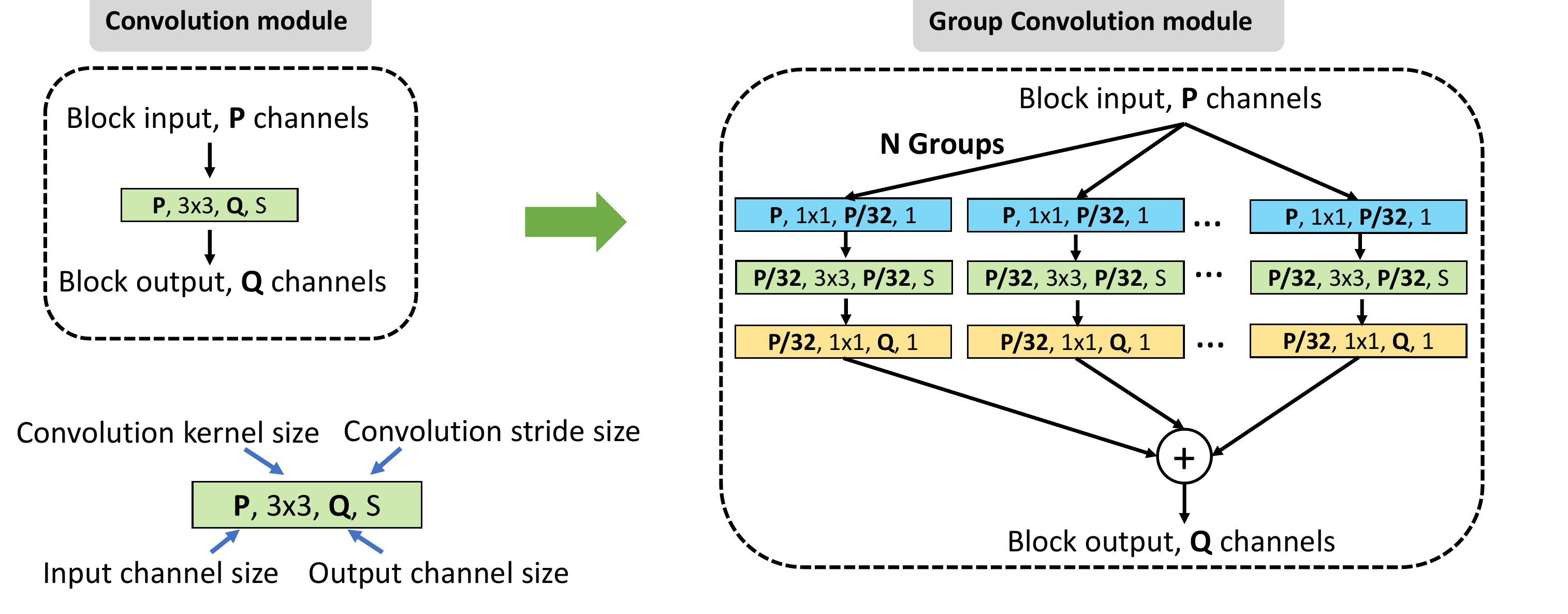}}
\caption{The schematic diagram of the {\txtc{group convolution module. The ReLU and Batch normalization layers are added after each convolution operation during implementation.} }}
\label{fig:group-conv-module}
\end{figure}
\subsubsection{Single input}
In many cases, only one MR sequence is available for attenuation correction, either T1-weighted, Dixon or UTE/ZTE. The network implemented for this scenario is based on the U-net structure \cite{ronneberger2015u}. The overall network architecture is summarized in Fig.~\ref{fig:u-net-structure}. It consists of repetitive applications of 1) $3 \times 3$ convolutional layer, 2) batch normalization layer, 3) ReLU layer, 4) convolutional layer with stride 2 for down-sampling, 5) transposed convolutional layer with stride 2 for up-sampling, and 6) mapping layer that concatenates the left-side features to the right-side. The input has nine channels with a spatial size of $144 \times 144$ and the bottom layer has an spatial size of $9 \times 9$. {\txtc{The number of features $N$ after the first convolution module is 16.}} To make full use of the axial information, nine neighboring axial slices were stacked occupying nine input channels to reduce the axial aliasing artifacts. As only Dixon images are utilized as single input, this method is referred as Dixon-Unet. 

\subsubsection{Multiple inputs}
For current PET/MR scanners, more than one MR sequence can be acquired for attenuation correction. For example, both Dixon and ZTE MR images are available in GE SIGNA scanner.  {\txtc{When multiple MR images are included as network input, the number of features $N$ after the first convolution module should be enlarged to digest the additional spatial information. For the U-net structure, the number of trainable parameters increases quadratically with $N$, and overfitting can be a serious pitfall when increasing the network complexity while not providing enough training pairs. It is shown in previous studies that designing a ``wider'' network can make more efficient use of model parameters \cite{szegedy2016rethinking,chollet2016xception,Xie_2017_CVPR}. To preserve the network capacity while restricting the network complexity, the group convolution module as illustrated in Fig.~\ref{fig:group-conv-module} was adopted to replace the convolution module when the network goes deeper. The group convolution module is similar to the module presented in ResNeXt network structure \cite{Xie_2017_CVPR}. Traditionally the convolution kernel considers cross-channel correlations and spatial correlations together. The group convolutional module presented in Fig.~\ref{fig:group-conv-module} first deals with the cross-channel correlation through 1x1 convolution and then handles the spatial correlation in smaller groups. The hypothesis is that when the network goes deeper, the spatial content and the cross-channel correlations can be decoupled \cite{chollet2016xception}.
In our implementation, $N$ is set to be 19 for the U-net with both Dixon and ZTE as input. For GroupU-net, the number of groups is set to be $N$ and we only use the group convolution module when the input channel size is $\ge 4N$. $N$ is set to $32$ to match with the number of trainable parameters in U-net (2.7 million). These two methods are labeled as DixonZTE-Unet and DixonZTE-GroupUnet, respectively.}}

\section{Experimental evaluations}
\subsection{Data sets }
The patient study was approved by the Institutional Review Board and all patients signed an informed consent before the examinations. In total forty patients acquired from 2014 to 2016 were used in this study. All patients had whole-body PET/CT, followed by additional PET/MRI scanning without second tracer administration. For both PET/CT and PET/MR, only data acquired in the bed position that includes the head are used in the study. {\txtc{No pathology in the brain was reported for any of the patients}}. The average patient weight was 73.2241 $\pm$ 17.0 kg (range, 39.5-109.8 kg). For PET/MRI, the average scan duration of the whole brain was 224.6 $\pm$ 133.7 s (range, 135-900 s). All forty patient data sets have Dixon MR images and fourteen patient data sets with additional acquired ZTE MR images.  Thirty seven of the total forty patient data sets had FDG scans. The average administered dose of FDG was 305.2 $\pm$ 73.9 MBq (range, 170.2-468.1 MBq). Twelve of the fourteen patients with additional ZTE scans had FDG PET scans.

PET/CT examinations were performed in the GE Discovery PET/CT scanner or the Siemens Biograph HiRez 16 PET/CT scanner. For CT images acquired from the GE Discovery PET/CT scanner, the reconstruction has a axial field of view (FOV) of $700$ mm and the matrix size is $512 \times 512$ with voxel size $2.73 \times 2.73 \times 3.75 \mbox{mm}^3$. For CT images acquired from the Siemens Biograph HiRez 16 PET/CT system, the reconstruction has a axial FOV of $500$ mm and the matrix size is $512 \times 512$ with voxel size $1.95 \times 1.95 \times 5.00 \mbox{mm}^3$. PET/MR examinations were performed in the GE SIGNA PET/MR system \cite{grant2016nema}. The transaxial and axial FOV of the PET/MR system is 600 mm and 250 mm, respectively. The crystal size is $4.0 \times 5.3 \times 25$ $\mbox{mm}^3$. PET images were reconstructed using the ordered subset expectation maximization (OSEM) algorithm with TOF information. The point spread function (PSF) \cite{alessio2010application} was also included to improve the image quality. Two iterations with sixteen subsets were run. The voxel size is $1\times 1 \times 2.87$ $\mbox{mm}^3$  and the image size is $300 \times 300 \times 89$. Dixon MR images were acquired using the head and neck coil array (repetition time, $\sim$4 ms; first echo time/second echo time, 1.3/2.6 ms; flip angle, 5$^{\circ}$; acquisition time, 18 s) and the image size is $256\times 256 \times 120$ with voxel size $1.93 \times 1.93 \times 2.6$ $\mbox{mm}^3$. ZTE images were acquired using the same head and neck coil array (repetition time, $\sim$0.7 ms; echo time, 0 ms; flip angle, 0.6$^{\circ}$; transmit/receive switching delay, 28 ms; readout duration, 440 ms; acquisition time, 41 s) and the reconstructed image size is $110\times 110 \times 110$ with voxel size $2.4 \times 2.4 \times 2.4$ $\mbox{mm}^3$ \cite{yang2017evaluation}. 
\subsection{Implementation details}
\begin{figure}[t]
\centering
\subfloat{\includegraphics[trim=4cm 11.5cm 4cm 11.3cm, clip, width=6.4in]{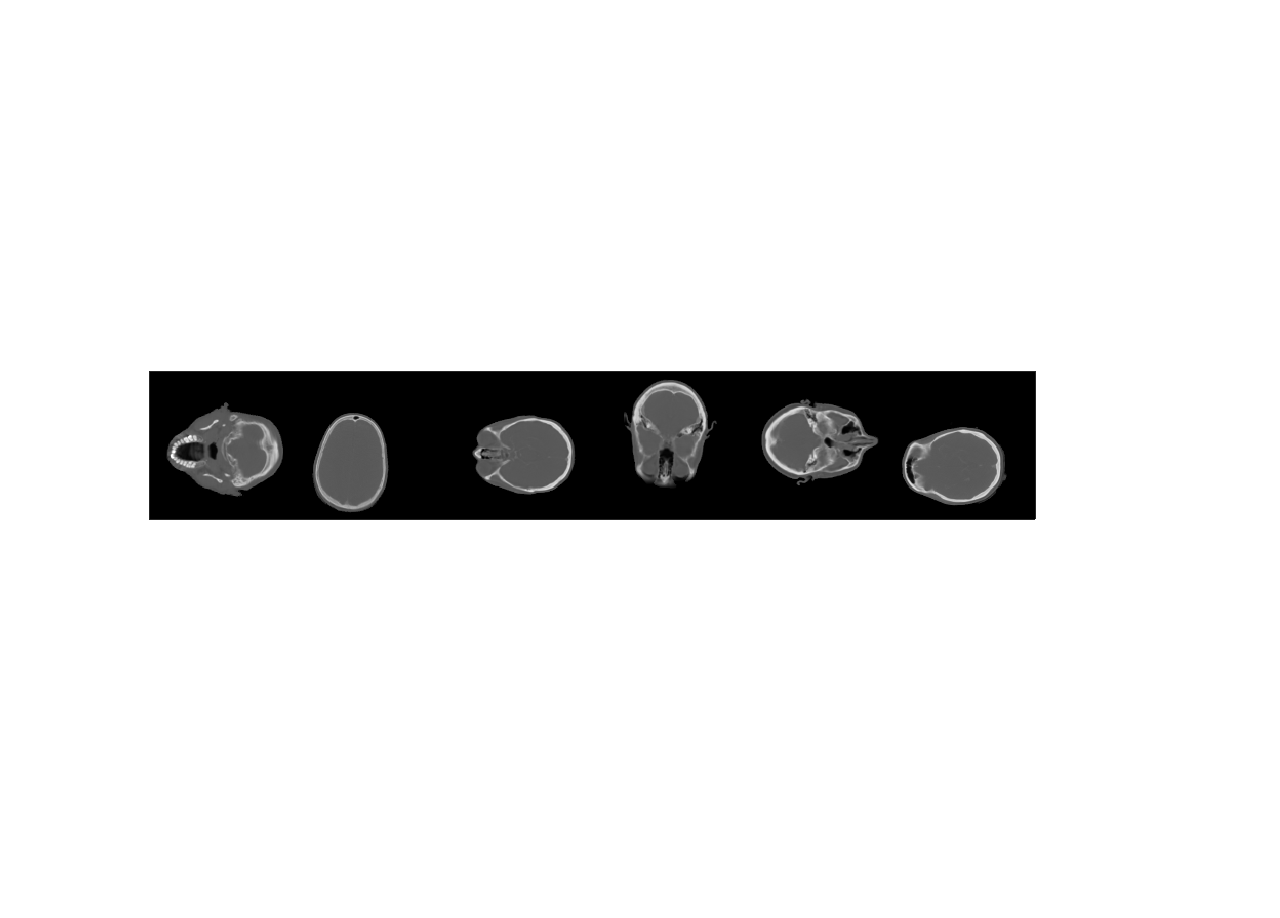}}\vspace{-0.5cm}\\
\subfloat{\includegraphics[trim=4cm 11.5cm 4cm 11.3cm, clip, width=6.4in]{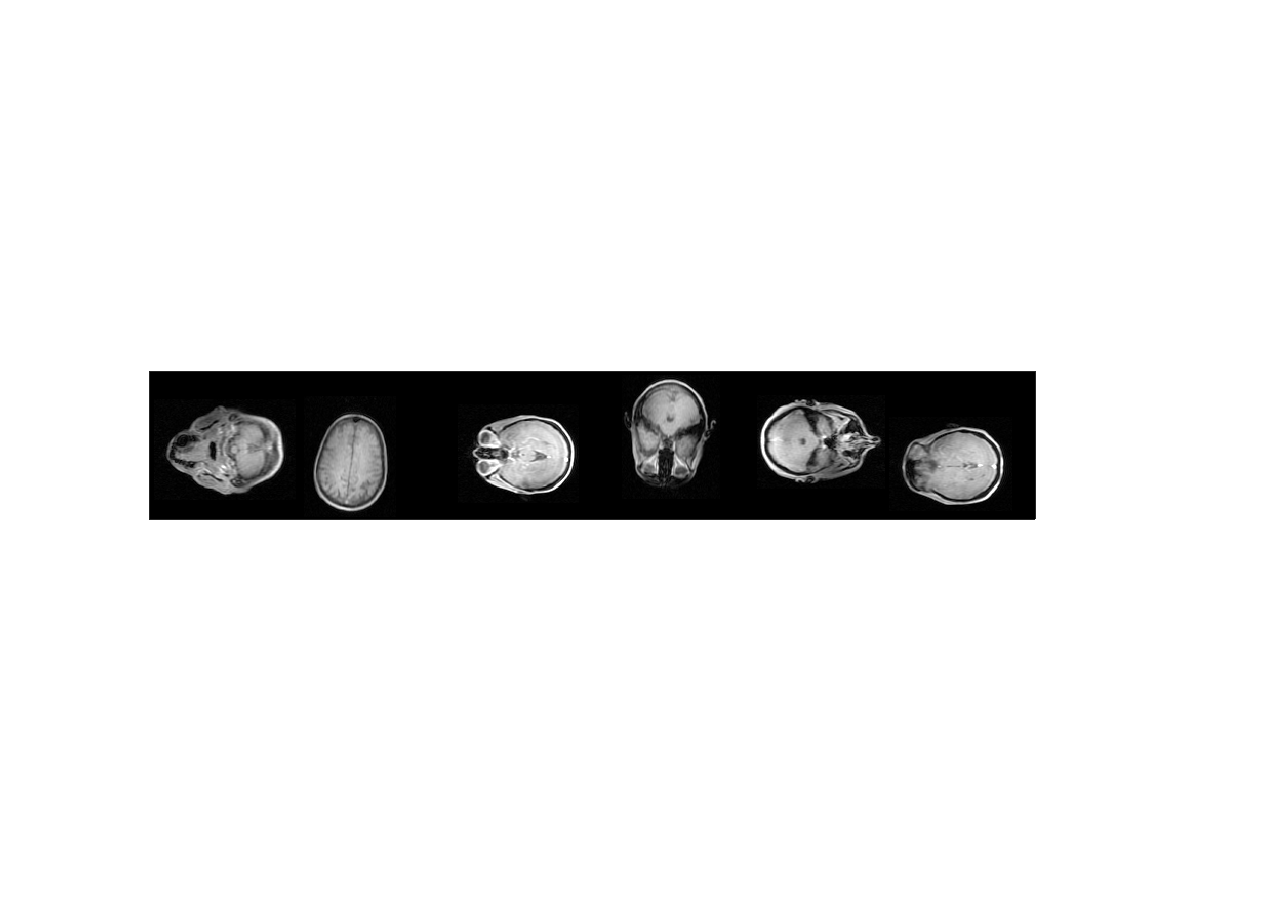}}\vspace{-0.5cm}\\
\subfloat{\includegraphics[trim=4cm 11.5cm 4cm 11.3cm, clip, width=6.4in]{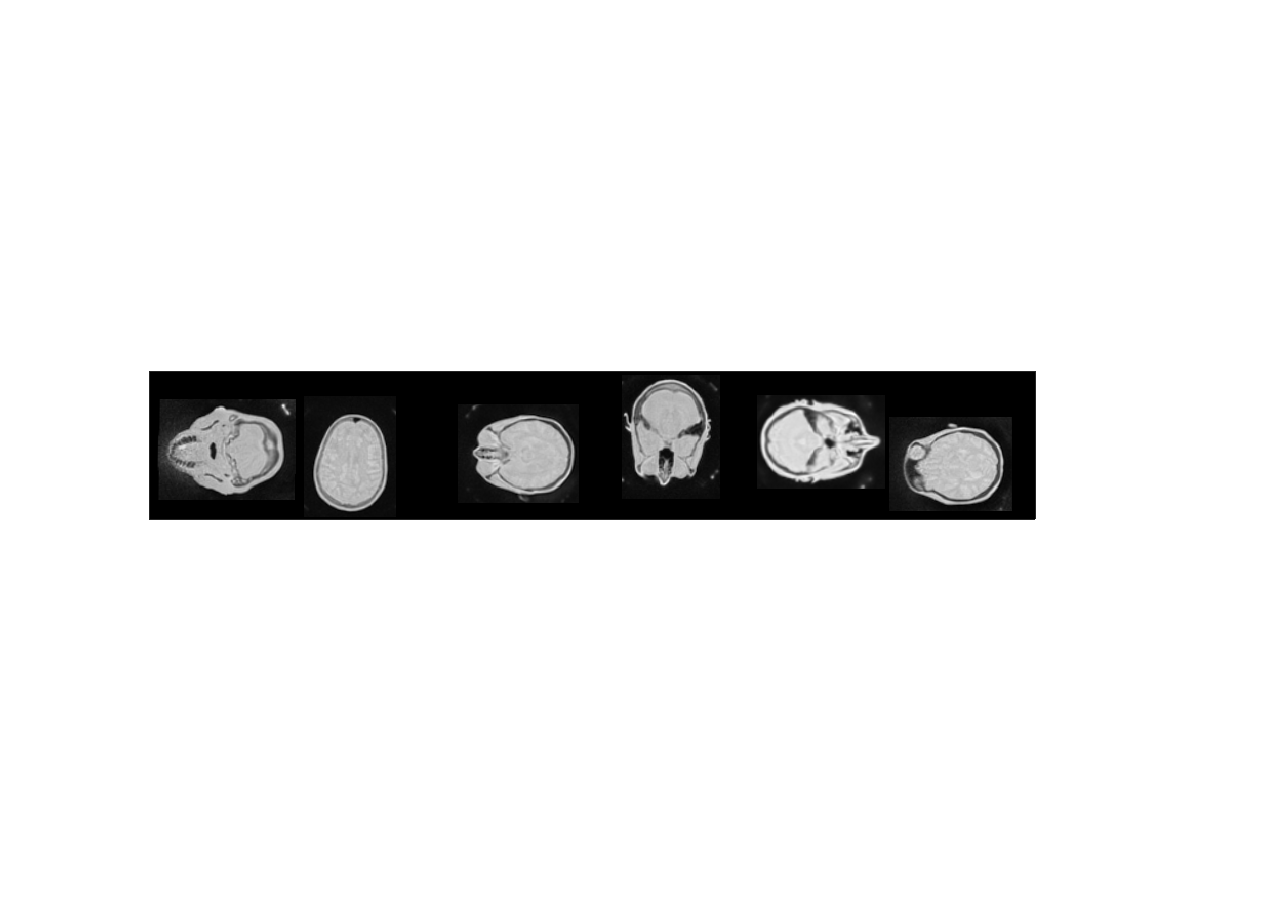}}
\caption{Examples of the training pairs used in the network training. Top row is the CT label image, middle and bottom rows are the corresponding Dixon MR images (middle) and ZTE MR images (bottom).}
\label{fig:show-training-pairs}
\end{figure}

When preparing the training pairs, we first registered CT images and ZTE images (if applicable) to the Dixon MR images through rigid transformation using the ANTs software \cite{avants2009advanced}. Then random rotation and permutation was performed on the training pairs to avoid over-fitting. Fig.~\ref{fig:show-training-pairs} shows some of the example pairs from different patient data sets used in the training phase for the multiple input scenario. When using only Dixon images as the input, in order to make full use of all the data sets in both the training and testing periods, the forty patient data sets were randomly separated into five groups. For each group, the whole eight data sets were used for testing and the remaining thirty two from other groups were employed in training. Among the forty patients, there are fourteen patients with additional ZTE scans. When using both Dixon and ZTE images as inputs, the fourteen patient data sets were randomly separated into seven groups. For each group the network was trained using the data sets from other groups. 

The network structures were implemented in TensorFlow using Adam algorithm as the optimizer \cite{kingma2014adam}.{\txtc{The learning rate and the decay rates used are the default settings in Tensorflow.}} For the single input case, the batch size was set to 60 and for the multiple input case, the batch size was set to 30. 1000 epochs were run for both cases as {\txtc{the training cost function becomes steady after 1000 epochs.}}

\subsection{References methods}
From the Dixon MR image, water and fat tissues were segmented and corresponding attenuation coefficients were assigned to generate the attenuation map. This method is labelled as Dixon-Seg. Alternatively, the patient MRI image can be registered to the MR template enabled by prior patients' MR and CT pairs through non-rigid registration. Air, soft tissue, sinus and bone exist in the generated CT image. This method is named as Dixon-Atlas \cite{wollenweber2013evaluation}. For the segmentation method using ZTE images, the ZTE images were first N4 bias corrected \cite{tustison2010n4itk} and then normalized by the median tissue value. Thresholding was performed to segment the images into air, soft tissue and bone regions. This method is labeled as ZTE-Seg. All of these three methods are available in the PET reconstruction tool box provided by the vendor.

\subsection{Evaluation metrics}
{\txtc{The predicted pseudo CT image quality was evaluated using the relative validation loss, defined as 
\begin{equation}
\mbox{Relative validation loss} = \frac{|\mbox{CT}_{\mbox{\small{pseudo}}} - \mbox{CT}_{\mbox{\small{true}}}|}{|\mbox{CT}_{\mbox{\small{true}}}|}, 
\end{equation}
where $\mbox{CT}_{\small{\mbox{pseudo}}}$ is the generated CT using different methods, and $\mbox{CT}_{\small{\mbox{true}}}$ denotes the ground-truth CT. Bone regions were also quantified using the Dice index, defined as
\begin{equation}
\mbox{Dice index} = 2\frac{\mbox{Bone}_{\mbox{\small{pseudo}}} \cap \mbox{Bone}_{\mbox{\small{true}}} }{\mbox{Bone}_{\mbox{\small{pseudo}}} + \mbox{Bone}_{\mbox{\small{true}}}}. 
\end{equation}
Regions with attenuation coefficient higher than 0.1083 $\mbox{cm}^{-1}$ (200 HU unit) were classified as the bone area. }} For PET image quantification, the relative PET error was used which is defined as
\begin{equation}
\mbox{Relative PET error} = \frac{|\mbox{PET}_{\mbox{\small{pseudoCT}}} - \mbox{PET}_{\mbox{\small{CT}}}|}{|\mbox{PET}_{\mbox{\small{CT}}}|}, 
\end{equation}
where $\mbox{PET}_{\mbox{\small{pseudoCT}}}$ is the PET image reconstructed using the pseudo CT, and $\mbox{PET}_{\mbox{\small{CT}}}$ is the PET image reconstructed using the ground-truth CT. The reason we use absolute value here is to ensure the total error will not vanish when summing up the voxel errors inside a region. As it is hard to visualize the error for all pixels, we calculated the relative PET error inside specific regions using the corresponding predefined masks. 
\subsubsection{Global quantification}
We performed a global brain quantification using the brain mask from MNI-ICBM
152 nonlinear 2009 version \cite{fonov2009unbiased}. The Dixon image of each patient was first registered to the MNI template. Then the MNI template was back-warped to the Dixon image space. Besides, a mask is defined to include the pixels whose intensity is larger than 30 percent of the max PET intensity \cite{ladefoged2017multi}. The final global brain mask is defined as the intersection of these two masks. Besides, the histograms of the error image inside the global brain mask, defined as $\mbox{PET}_{\mbox{\small{pseudoCT}}} - \mbox{PET}_{\mbox{\small{CT}}}$, were calculated to compare the global performance regarding the bias and standard deviation. 
\subsubsection{Regional quantification}
Apart from the whole brain quantification, we are also interested in the regional brain quantifications as they each play crucial roles in specific neuroscience studies. The automated anatomical labeling (AAL) template \cite{holmes1998enhancement} was back-warped to the PET image space and defined the regions. Four cortex lobes as well as the inner deep regions were used in the quantification. The mean and standard deviation of the relative PET error across all patients for each of the methods were calculated for all the regions and the whole brain.

\section{Results}
\subsection{Using Dixon MR images as input}

\begin{figure}[b]
\centering
\subfloat{\includegraphics[trim=4cm 8.3cm 2cm 7.5cm, clip, width=5.2in]{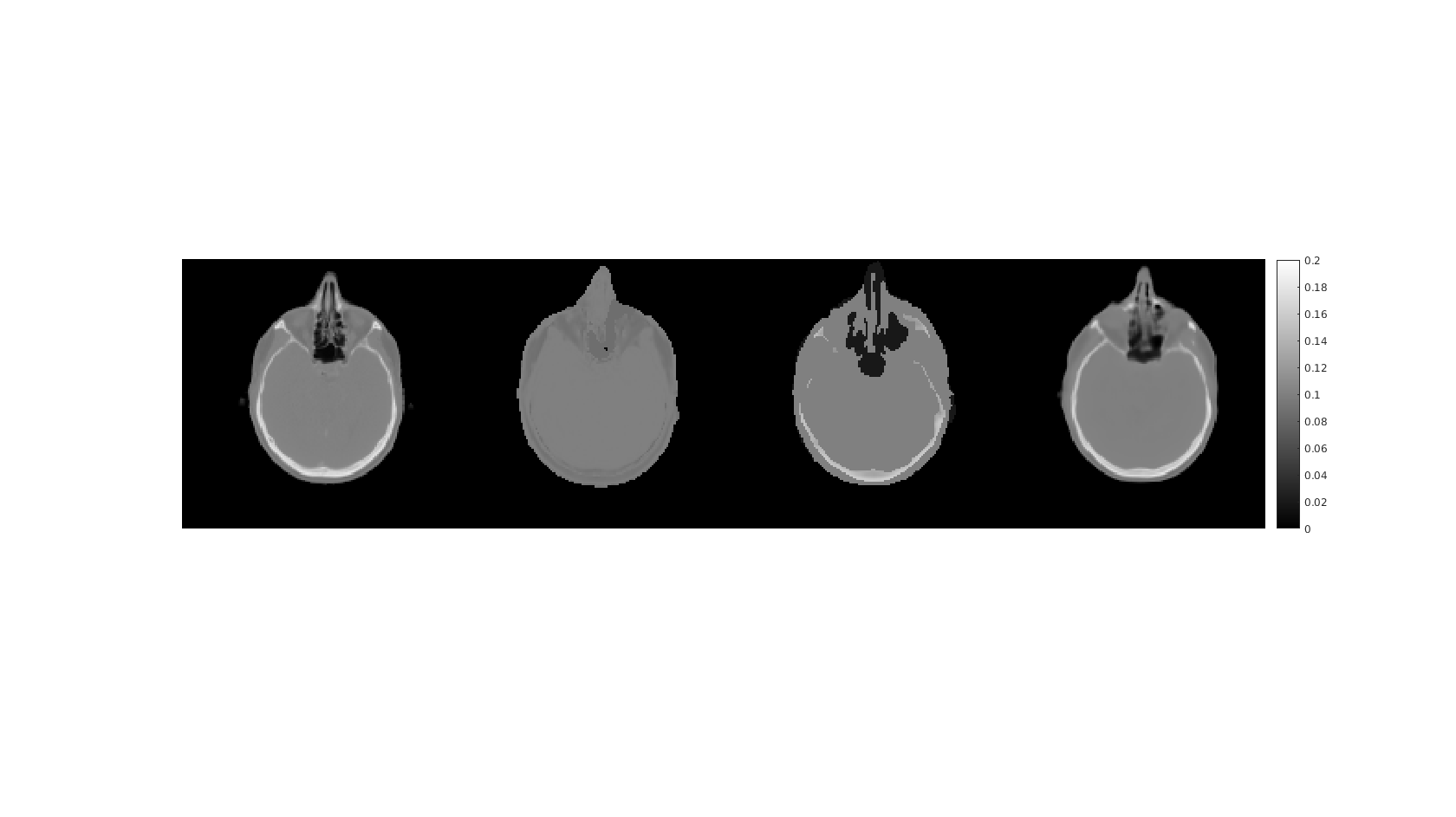}}\vspace{-0.5cm}\\
\subfloat{\includegraphics[trim=4cm 8.3cm 2cm 7.5cm, clip, width=5.2in]{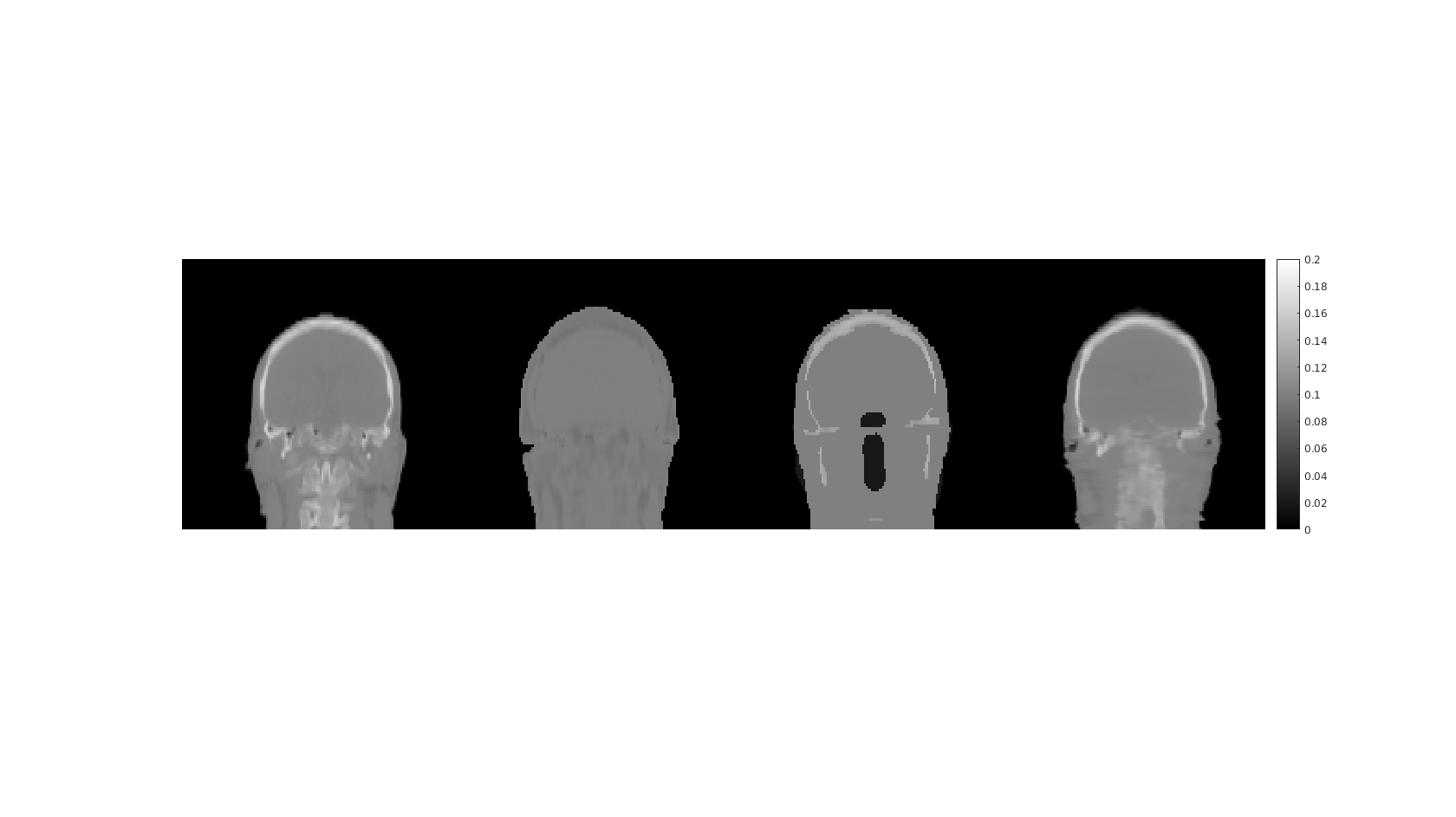}}\vspace{-0.5cm}\\
\subfloat{\includegraphics[trim=4cm 8.3cm 2cm 7.5cm, clip, width=5.2in]{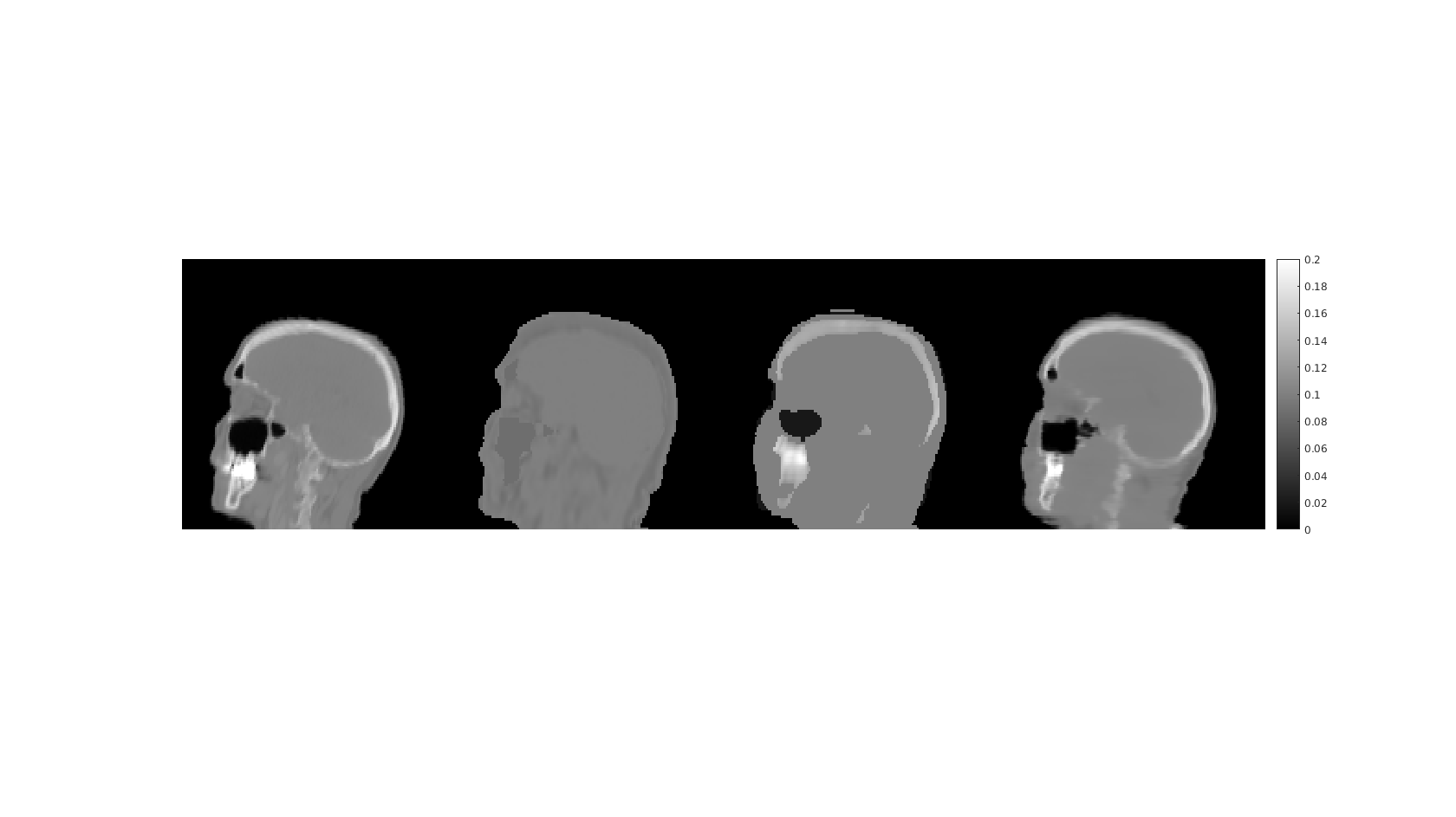}}\\
\caption{Three views of {\txtc{the attenuation maps (unit, cm${^{-1}}$) derived from}} the true CT image (first column) and the generated pseudo CT images using the Dixon-Seg method (second column), Dixon-atlas method (third column) and the proposed Dixon-Unet method (last column).}
\label{fig:dixon-ct-compare}	
\end{figure}

\begin{table}[t]
\caption{\txtc{Comparisons of the generated pseudo-CT images when only Dixon images are available (based on 40 patient data sets).} The Dice index of bone regions was computed for the whole brain, regions above and below the eyes.}
\label{tab:ct-dixon-compare}
\centering
\begin{tabular}{c|c|c|c|c}
    \hline \hline 
 	 Methods &  \makecell{Relative validation \\loss ($\%$)} &\makecell{ Dice of bone \\ whole} & \makecell{Dice of bone \\ above eye}& \makecell{Dice of bone \\ below eye} \\
 	\hline
    Dixon-Seg &  $32.70 \pm 5.38$ & $-$ & $-$  & $-$\\
	
    \hline
    Dixon-Atlas &  $22.86 \pm 2.34$ & $0.52 \pm 0.05$ & $0.61 \pm 0.06$  & $0.30 \pm 0.05$\\
	\hline
    Dixon-Unet & $\bb{13.84} \pm 1.43$ & $\bb{0.76} \pm 0.04$ & $\bb{0.82} \pm 0.04$ & $\bb{0.63} \pm 0.06$\\
    \hline \hline
\end{tabular}
\end{table}

\begin{figure}[h]
\centering
\subfloat{\includegraphics[trim=4cm 8.1cm 2cm 7.2cm, clip, width=5in]{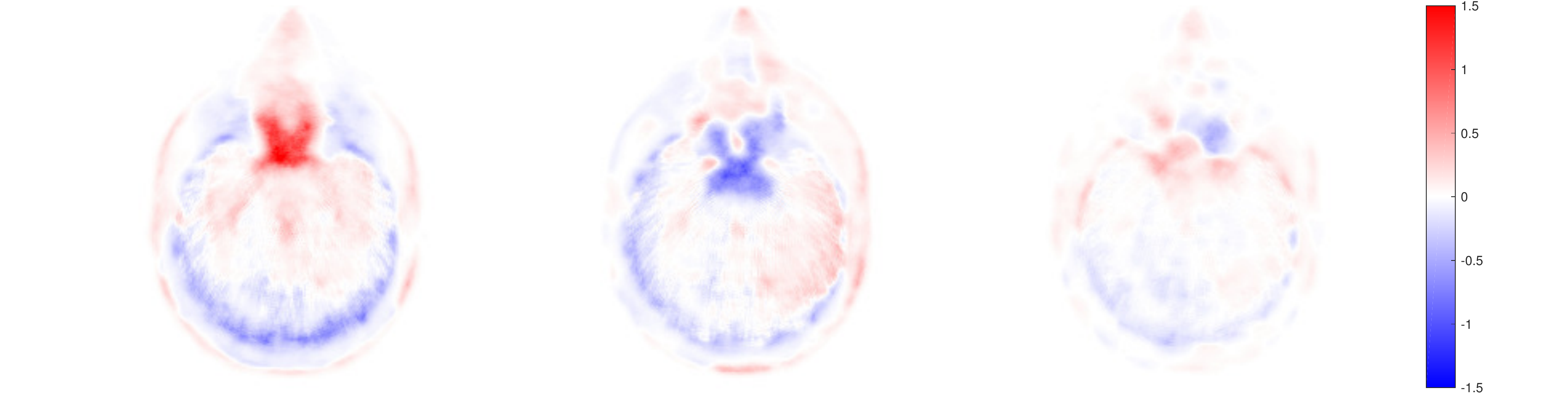}}\\
\subfloat{\includegraphics[trim=4cm 8.3cm 2cm 7.5cm, clip, width=5in]{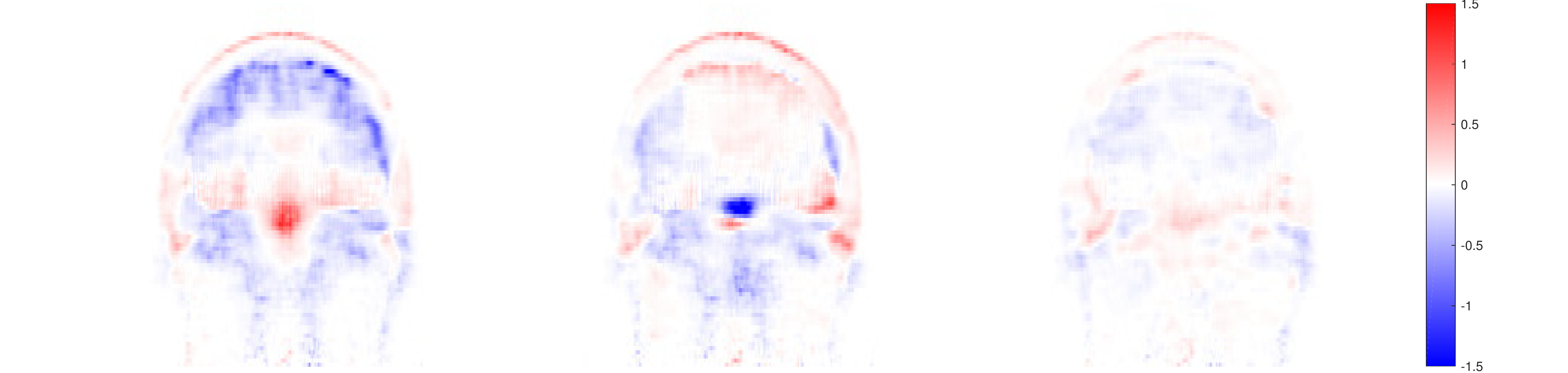}}\\
\subfloat{\includegraphics[trim=4cm 8.3cm 2cm 7.5cm, clip, width=5in]{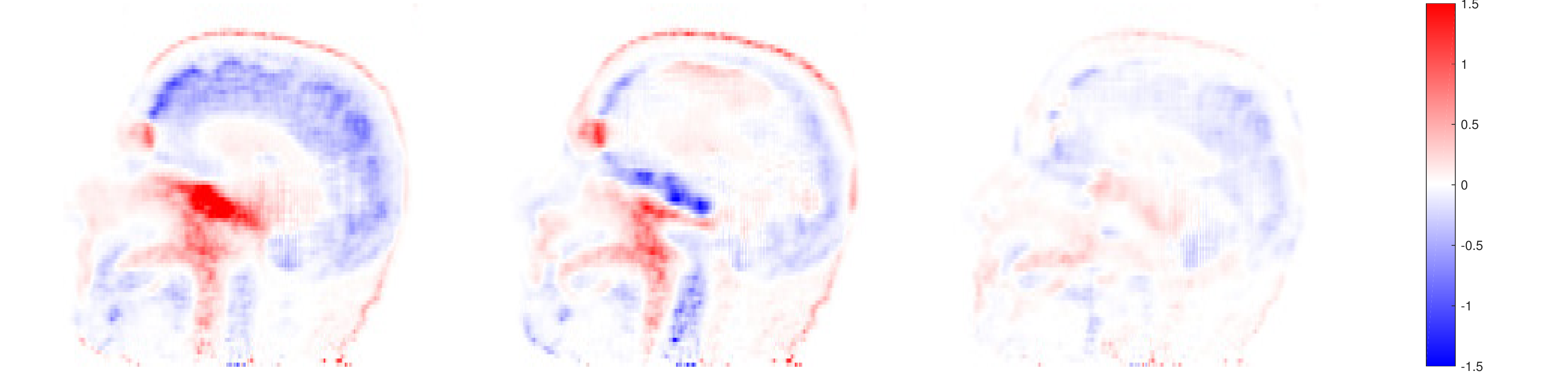}}\\
\caption{Three views of the PET reconstruction error images {\txtc{\small{($\mbox{PET}_{\mbox{\small{pseudoCT}}} - \mbox{PET}_{\mbox{\small{CT}}}$, unit: SUV)}}} using the Dixon-Seg method (left column), the Dixon-atlas method (middle column) and the proposed Dixon-Unet method (right column).}
\label{fig:pet-error-dixon}
\end{figure}

We first performed a comparison of the proposed Dixon-Unet method with the Dixon-Seg and Dixon-Atlas methods using all data sets. Fig.~\ref{fig:dixon-ct-compare} shows three orthogonal views of the ground truth CT images and the generated pseudo CT images using different Dixon-based methods for one patient. Compared with the atlas method, the CT image produced by the proposed Dixon-Unet method has better bone and sinus structures. The Dixon-Seg method only shows the water and fat tissues. {\txtc{Table.~\ref{tab:ct-dixon-compare} presents the quantitative comparison of the predicted CT images using relative validation loss and the Dice index. Clearly the Dixon-Unet method has the smallest validation loss and the highest Dice index in the bone region. }}Fig.~\ref{fig:pet-error-dixon} presents the PET reconstruction error images using the attenuation map produced from the pseudo CT images shown in Fig.~\ref{fig:dixon-ct-compare}. Evidently the Dixon-Seg method has the largest error, especially near the bone and air-cavity regions. The Dixon-Atlas method produces smaller errors compared with the Dixon-Seg method, but still has large errors near the bone and the air cavity. Compared with these two methods, Dixon-Unet method shows smaller errors for the whole brain.

\begin{figure}[h]
\centering
\subfloat{\includegraphics[trim=0cm 0cm 0cm 0cm, clip, width=4.3in]{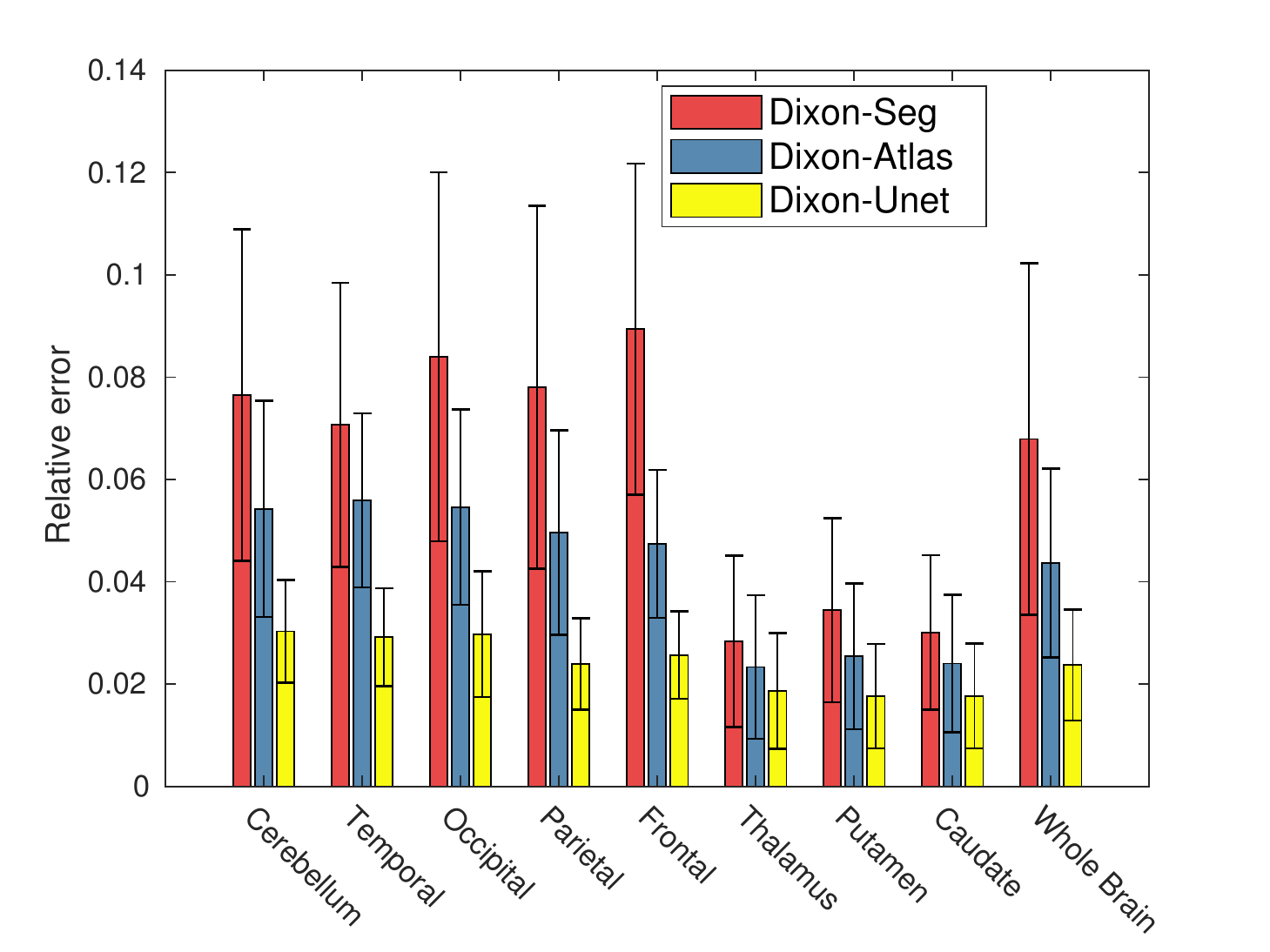}}
\caption{The bar plot of the mean relative PET error for all the patient data sets. Standard deviation of the relative PET error for all the patients are plotted as the error bar. }
\label{fig:dixon-region-barplot}
\end{figure}
\begin{figure}[h]
\centering
\subfloat{\includegraphics[trim=0cm 0cm 0cm 0cm, clip, width=4.3in]{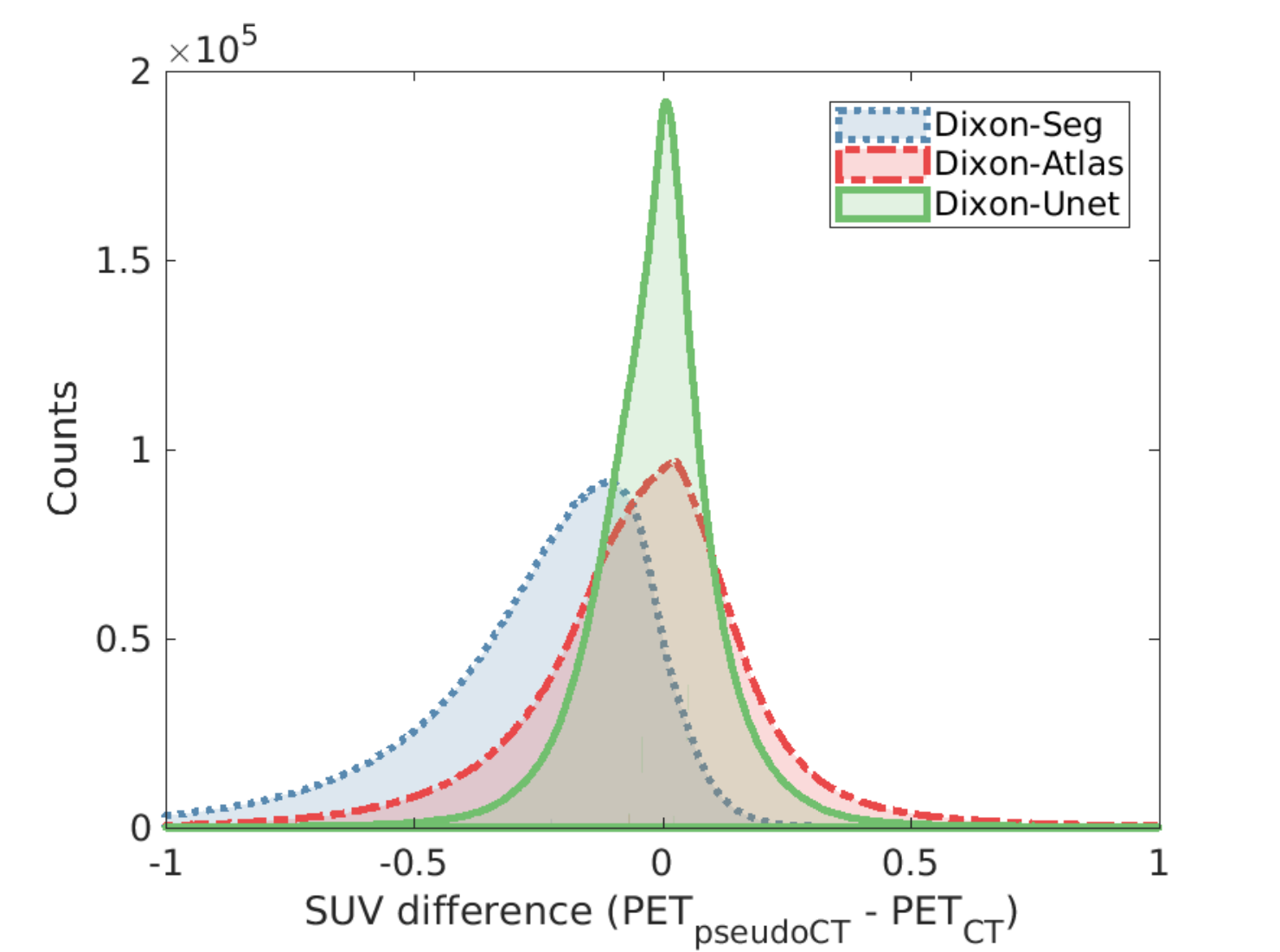}}
\caption{The histogram of PET error images for the three Dixon methods.}
\label{fig:dixon-peterror-hist}
\end{figure}

To quantitatively characterize the influence of different attenuation correction methods on PET images, the mean relative PET error across all the data sets for the whole brain and different regions were calculated and presented in Fig.~\ref{fig:dixon-region-barplot}. Clearly in all regions the Dixon-Unet method is the best among all Dixon methods. The Dixon-Seg method has the largest error due to the missing of bone signals. Comparing the standard deviations, Dixon-Unet method has the smallest standard deviation in all regions, meaning it is robust across different populations by using the information from other patient data sets. For all regions, the error of the Dixon-Unet method is below $3\%$. Fig.~\ref{fig:dixon-peterror-hist} shows the histogram plot of the PET error images for the three methods. The plot indicates that the error image of the Dixon-Unet method has the smallest standard deviation and the histogram shape is more like a Gaussian distribution with zero mean. The histogram shapes of the Dixon-Seg and Dixon-Atlas methods are more screwed. {\txtc{Specially the Dixon-Seg method is negatively biased due to missing bone.}}

\subsection{Using both Dixon and ZTE MR images as input}

\begin{figure}[t]
\centering
\subfloat{\includegraphics[trim=5cm 8.9cm 2cm 8.6cm, clip, width=6in]{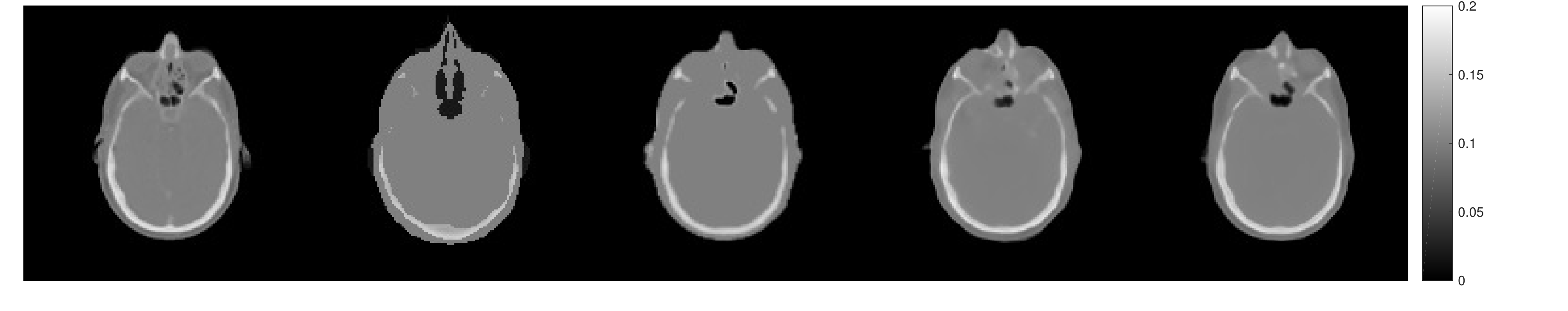}}\vspace{-0.6cm}\\
\subfloat{\includegraphics[trim=5cm 8.9cm 2cm 8.6cm, clip, width=6in]{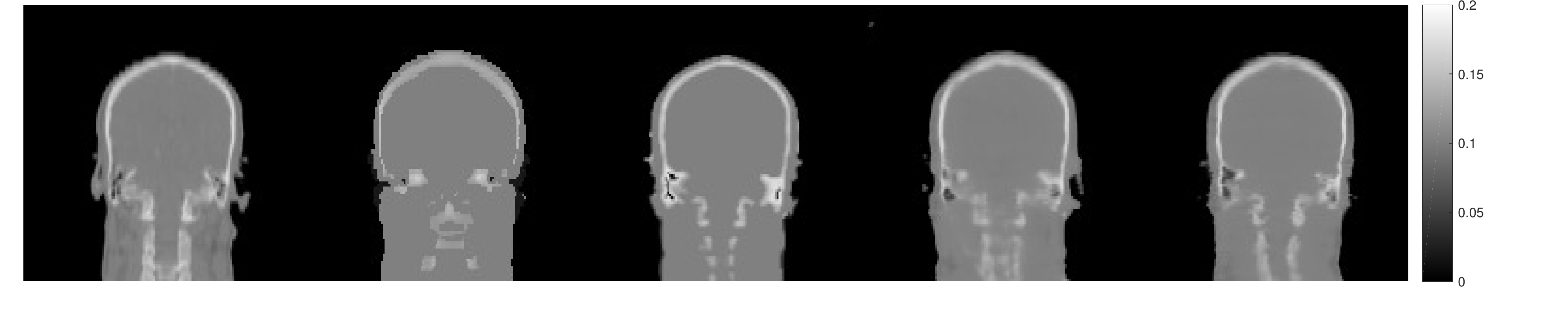}}\vspace{-0.6cm}\\
\subfloat{\includegraphics[trim=5cm 9.5cm 2cm 8.6cm, clip, width=6in]{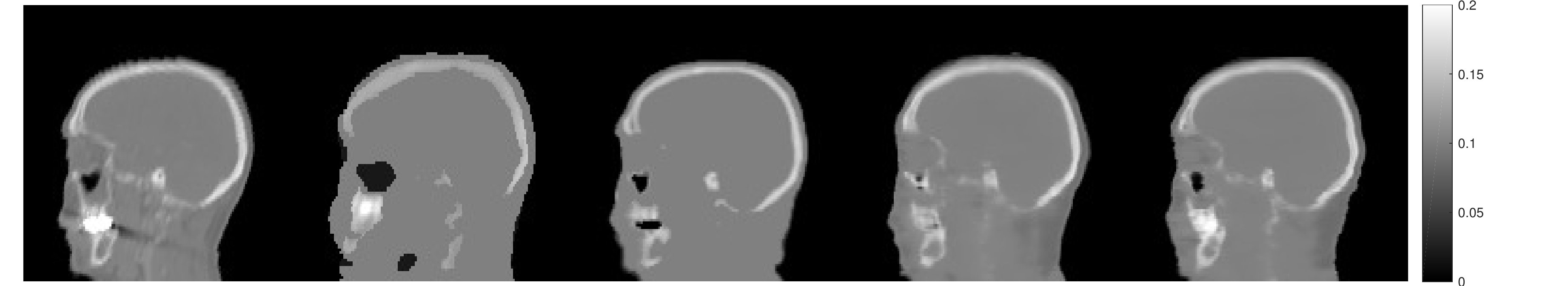}}\\
\caption{{\txtc{Comparison of the true CT image (first column) with generated pseudo CT images using the Dixon-Atlas method (second column), the ZTE-Seg method (third column), the DixonZTE-Unet method (fourth column) and the DixonZTE-GroupUnet method (last column).}}}
\label{fig:zte-ct-show}
\end{figure}

\begin{figure}[t]
\centering
\subfloat{\includegraphics[trim=0cm 0cm 0cm 0cm, clip, width=3.5in]{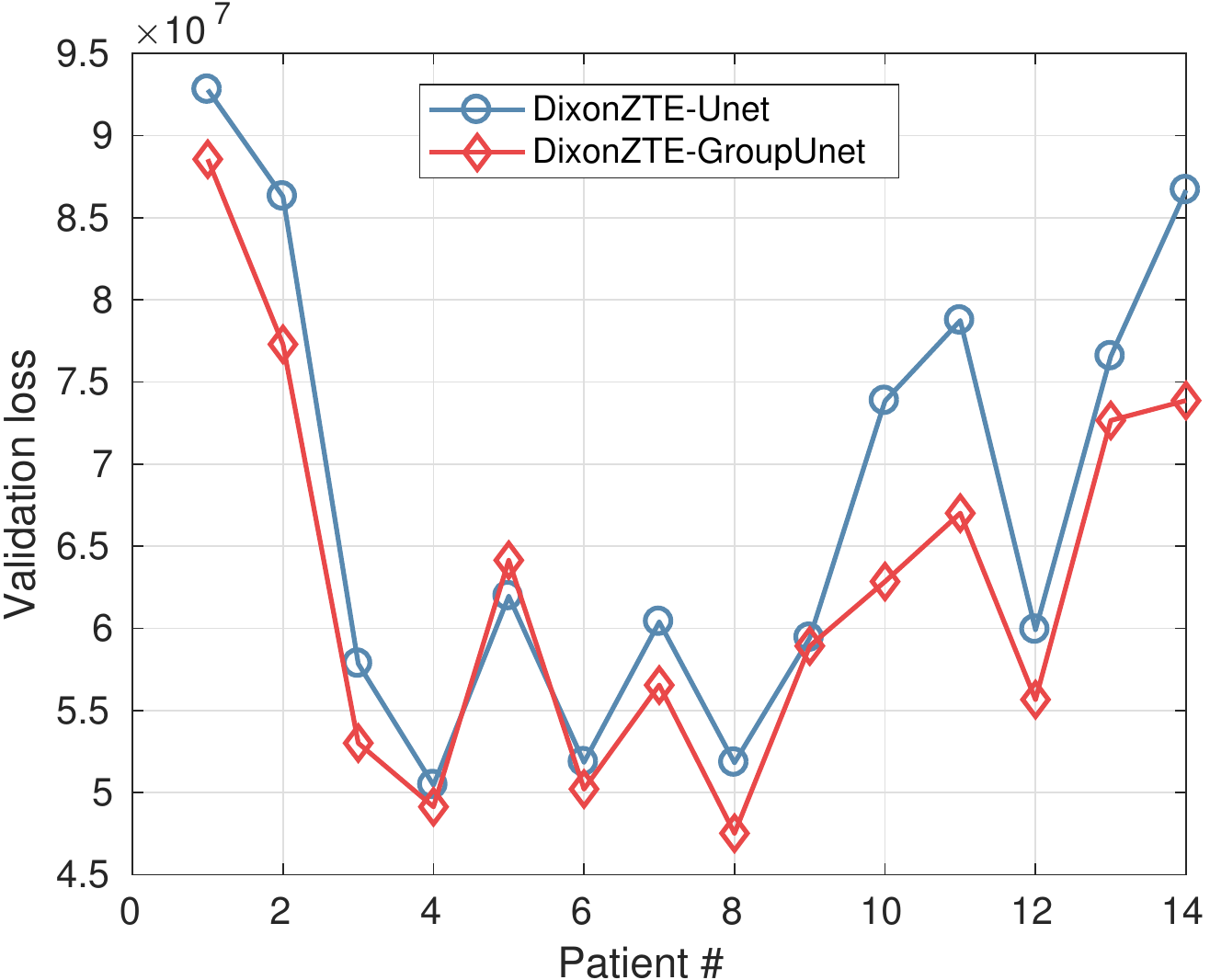}}
\caption{\txtc{Comparison of the validation loss regarding the predicted CT images using U-net and the proposed GroupU-net when both Dixon and ZTE MR images are used as network input.}}
\label{fig:ct-validation-loss}
\end{figure}
\begin{table}[t]
\caption{\txtc{The comparison of the generated pseudo-CT when both Dixon and ZTE images are available (based on 14 patient data sets). The Dice index of bone was computed for the whole brain, regions above and below the eyes.} }
\label{tab:kinetics}
\centering
\begin{tabular}{c|c|c|c|c}
    \hline \hline 
 	 Methods &  \makecell{Relative validation \\loss ($\%$)} &\makecell{ Dice of bone \\ whole} & \makecell{Dice of bone \\ above eye}& \makecell{Dice of bone \\ below eye} \\
    \hline
    Dixon-Atlas &  $23.33 \pm 3.23$ & $0.52 \pm 0.05$ & $0.61 \pm 0.06$  & $0.29 \pm 0.05$\\
	\hline
   	ZTE-Seg &  $16.20 \pm 2.28$ & $0.69 \pm 0.05$ & $0.75 \pm 0.05$ & $0.56 \pm 0.07$\\
	\hline
    DixonZTE-Unet & $13.58 \pm 1.53$ & $0.77 \pm 0.04$ & $0.83 \pm 0.04$ & $0.66 \pm 0.07$\\
	\hline
	DixonZTE-GroupUnet &  $\bb{12.62} \pm 1.46$ & $\bb{0.80} \pm 0.04$  & $\bb{0.86} \pm 0.03$ & $\bb{0.69} \pm 0.06$ \\
    \hline \hline
\end{tabular}
\end{table}

\begin{figure}[h]
\centering
\subfloat{\includegraphics[trim=5cm 8.9cm 2cm 7.6cm, clip, width=5.8in]{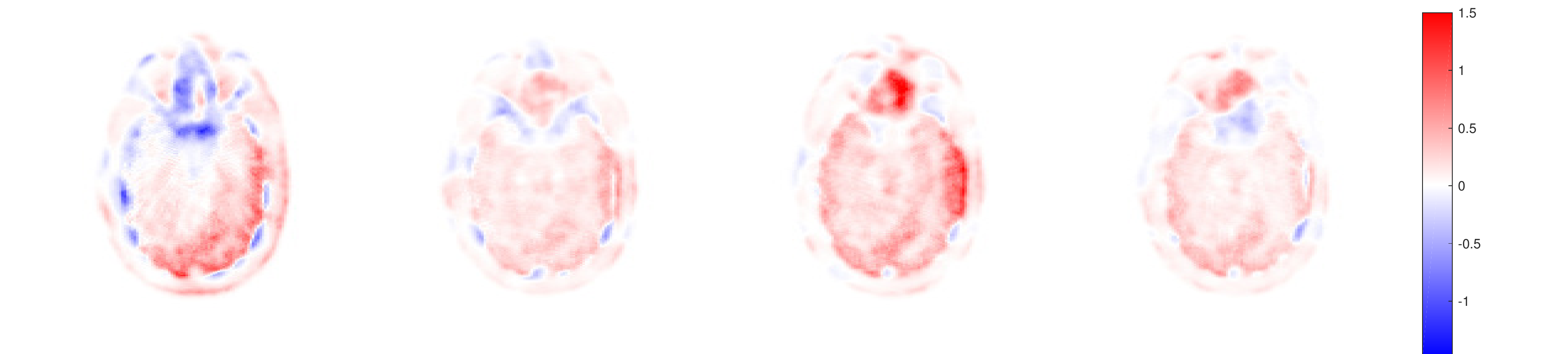}}\vspace{-0.8cm}\\
\subfloat{\includegraphics[trim=5cm 8.9cm 2cm 8.3cm, clip, width=5.8in]{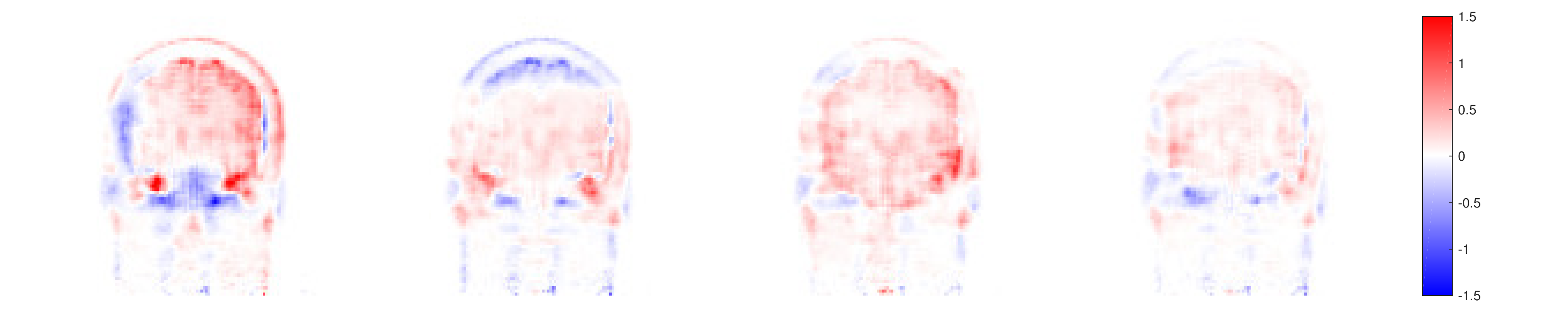}}\vspace{-0.8cm}\\
\subfloat{\includegraphics[trim=5cm 9.4cm 2cm 8.3cm, clip, width=5.8in]{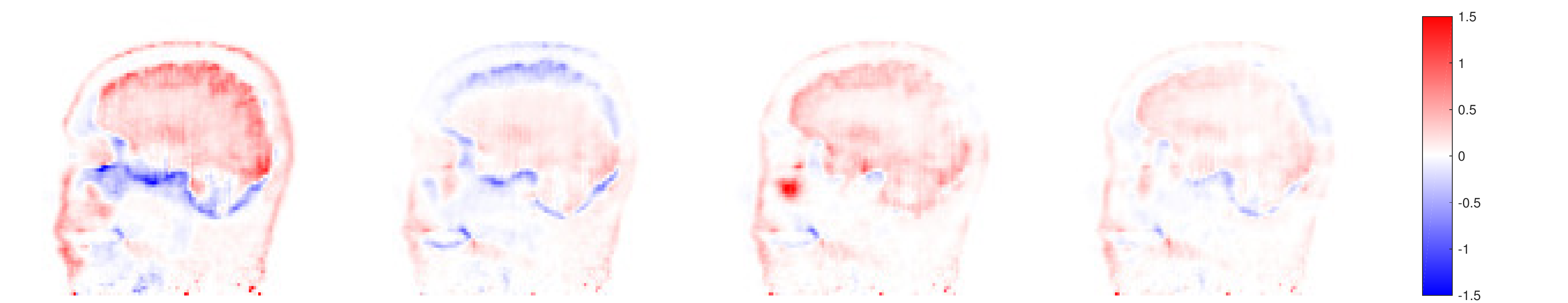}}\\
\caption{\txtc{\small{PET reconstruction error images {\txtc{\small{($\mbox{PET}_{\mbox{\small{pseudoCT}}} - \mbox{PET}_{\mbox{\small{CT}}}$, unit: SUV)}}} using the Dixon-Atlas method (first column), the ZTE-Seg method (second column),the DixonZTE-Unet method (third column) and the DixonZTE-GroupUnet method (last column). } }}
\label{fig:zte-pet-recon}
\end{figure}

\begin{figure}[h]
\centering
\subfloat{\includegraphics[trim=0cm 0cm 0cm 0cm, clip, width=4.3in]{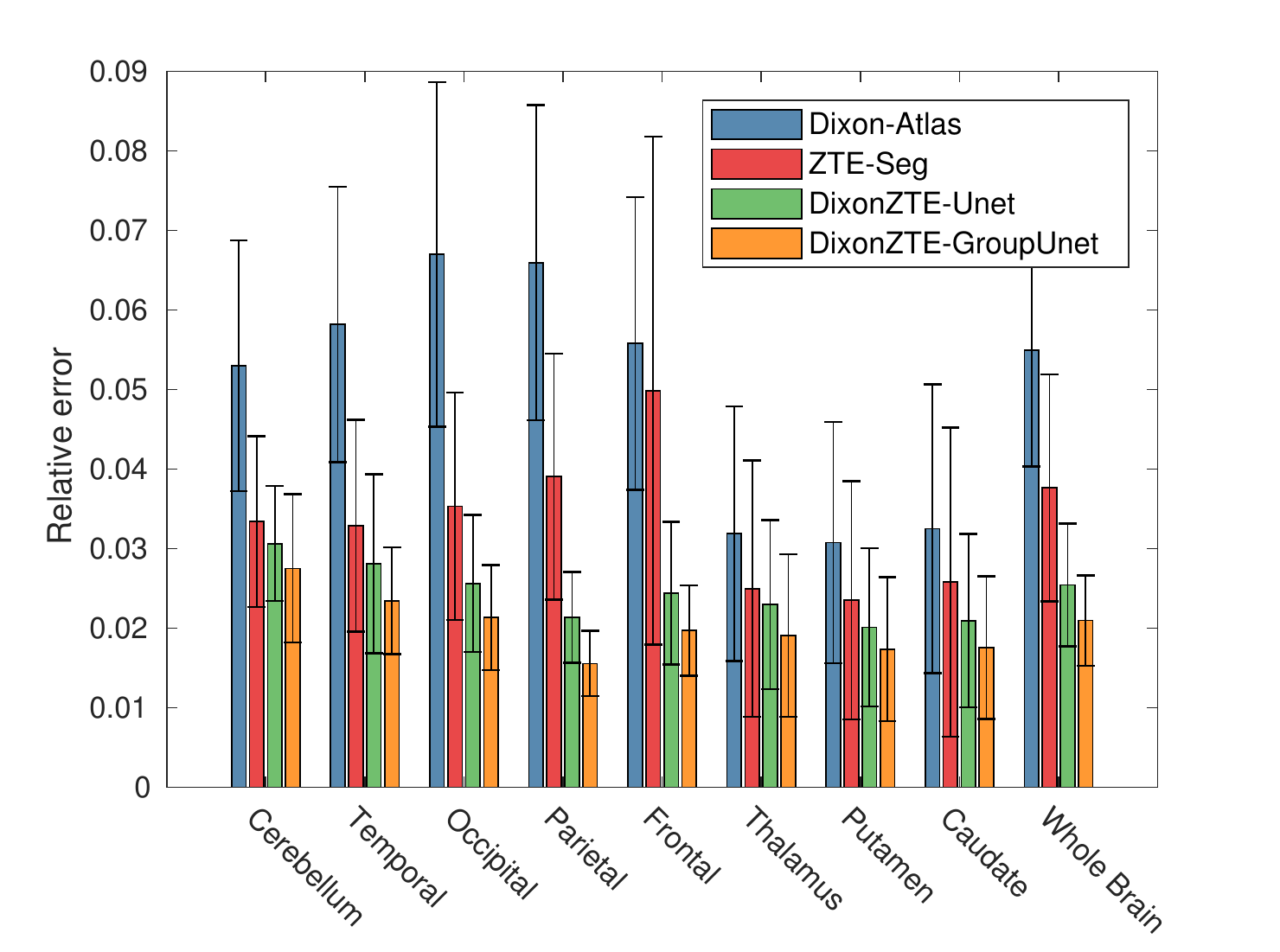}}
\caption{\txtc{\small{The bar plot of the mean relative PET error for the patient data sets with both Dixon and ZTE images. Standard deviations of the absolute error for all the patients are plotted as the error bar.}}}
\label{fig:zte-peterror-hist}
\end{figure}

\begin{figure}[t]
\centering
\subfloat{\includegraphics[trim=0cm 0cm 0cm 0cm, clip, width=4.3in]{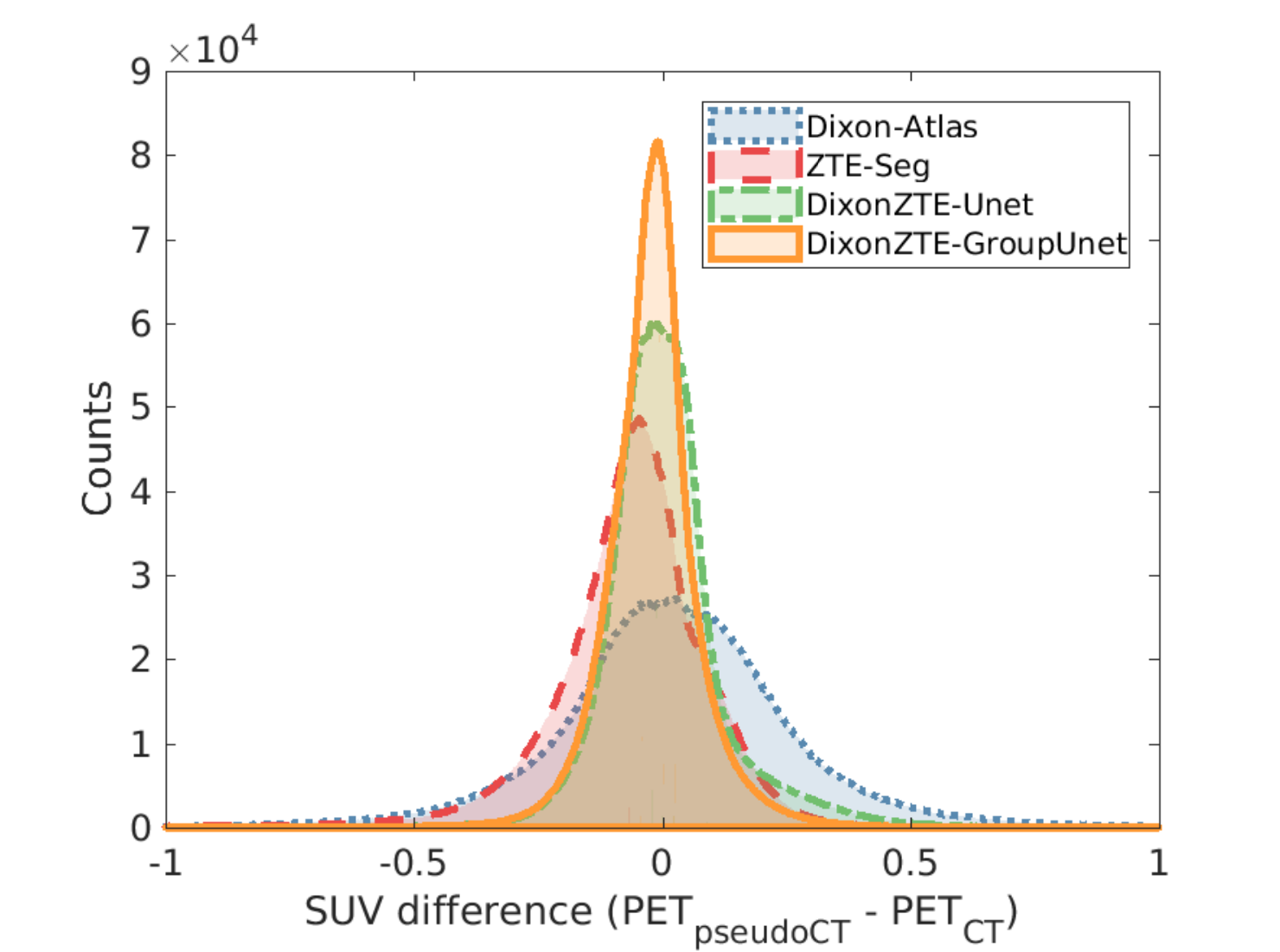}}
\caption{\txtc{\small{The histogram plot of the {PET SUV difference} inside the whole brain for Dixon-Atlas, ZTE-Seg, DixonZTE-Unet and DixonZTE-GroupUnet methods.}} }
\label{fig:zte-histogram}
\end{figure}

In the following analysis, results using Dixon-Atlas, ZTE-Seg, DixonZTE-Unet and DixonZTE-GroupUnet methods were presented and compared using the twelve patient data sets with FDG scans. Fig.~\ref{fig:zte-ct-show} shows three orthogonal views of the ground truth CT images as well as the generated pseudo CT images using different methods for one patient. Compared to the Dixon-Atlas method, the ZTE-Seg method can recover most of the bone regions as the contrast between the bone and neighboring pixels is good in the ZTE MR image. The images generated using the neural network methods are generally similar to the images generated using the ZTE-Seg method, but with more details revealed and closer to the CT ground truth. {\txtc{To compare the pseudo-CT qualities for each data set, Fig.~\ref{fig:ct-validation-loss} shows the CT validation loss using the U-net and GroupU-net structures. The proposed GroupU-net method has lower validation loss in 13 out of 14 data sets.}} {\txtc{Table.~\ref{tab:kinetics} presents the quantitative comparison of the predicted CT images. The proposed GroupU-net method has the smallest validation loss and the highest Dice index in the bone region.}} Fig.~\ref{fig:zte-pet-recon} gives three views of the PET reconstruction error images based on the corresponding pseudo CT images presented in Fig.~\ref{fig:zte-ct-show}. The Dixon-Atlas method has the largest error and the DixonZTE-GroupUnet method has the smallest error. Fig.~\ref{fig:zte-peterror-hist} shows the plot of the mean relative PET error for all twelve patients across different regions. Clearly the Dixon-Atlas method has the largest mean error in all regions and the ZTE-Seg method generates smaller errors as compared with the Dixon-Atlas method. The proposed neural network methods can be better than both Dixon-Atlas and ZTE-Seg methods. Specially, the DixonZTE-GroupUnet can produce the smallest errors in all regions. This trend can also be observed in the histogram plot of the PET error images shown in Fig.\ref{fig:zte-histogram}. The DixonZTE-GroupUnet method has both the smallest standard deviation and the smallest systematic bias. 

\section{Discussion}
Dixon MR acquisition is simple and fast. It is widely deployed in current PET/MR systems as an option for further attenuation map derivation. As the signal intensity is low in the bone region, it is hard to segment the bone out. In this work, we employed the deep neural network method to predict pseudo CT images from Dixon images. From the CT images presented in Fig.~\ref{fig:dixon-ct-compare}, we can notice that the shape of the bone region predicted by the neural network method is much better than the atlas method. This indicates the neural network can recognize the bone region from the Dixon image input. Further quantitative analysis based on 40 patient data sets reveals that the mean relative PET error of the whole brain using the neural network method is within 3\%, which demonstrates the reproducibility of the proposed method.

With the developments of new MR sequences, multiple MR images are available during the same scan. {\txtc{It is thus crucial to find an optimum way integrating the information from multiple MR images while not increasing the network complexity much, especially when the training data sets are not large enough. In this work we have proposed a modified U-net structure, named GroupU-net, to digest both Dixon and ZTE information through group convolution modules when the network goes deeper. The group convolution module first considers the cross-channel correlation through $1\times 1$ convolution, and then handles the spatial correlation in smaller groups.}} Quantification analysis shows that the GroupU-net structure has better performance than the U-net structure when the network complexity is the same. {\txtc{This demonstrates that model parameters can be used more efficiently by making the network wider when the network goes deeper. }}It also shows that improving the network structure can generate better attenuation maps. Designing and testing different network structures will be one of our future work.

{\txtc{For the case of using both Dixon and ZTE images as network input, there are 12 patient data sets in each training group. Quantitative analysis demonstrates that 12 patient data sets can be used to train a network which provides higher CT prediction accuracy than the state-of-art methods. One limitation of this work is that no brain pathology was reported for the brain data sets employed in this study. We are unsure about the prediction accuracy for MR images with abnormal regions. If the test data do not lie in the training space due to population difference, the trained network may not accurately recover unseen structures. The robustness of the trained network to diseased data sets deserves further evaluations.}}

As for the objective function employed in the network training, L1 norm was found to be better than L2 norm. L2 norm results in blurrier images. We also tried another objective function by including additional L1 difference between the gradient images of the ground-truth CT and pseudo CT in both the horizontal and vertical directions. Though the generated CT image had a sharper bone, quantification for the inner regions, such as the putamen and caudate, showed worse results. {\txtc{The sizes of the air-cavity regions in the ZTE and Dixon MR images are different. As different methods extract information from Dixon only, ZTE only, or Dixon and ZTE combined, there will be difference about the delineations of air-cavity as shown in Fig.~\ref{fig:dixon-ct-compare} and Fig.~\ref{fig:zte-ct-show}. }} Additionally, we noticed that for the MR and CT images acquired in two different scanners, the jaw and head-neck regions could not be registered well in some cases due to position difference. This can produce errors as the training presumes that the CT and MR images match perfectly. Generalized adversarial networks which do not depend on the paired MR and CT images might help solve this problem.





\section{Conclusion}
We have proposed a neural network method to generate the continuous attenuation map for brain PET imaging based on the Dixon MR images only, and based on Dixon and ZTE images combined. Analysis using real data sets shows that the neural network method can produce smaller PET quantification errors as compared to other standard methods. When both Dixon and ZTE images are available, {\txtc{the proposed GroupU-net structure, which extracts features from Dixon and ZTE images through group convolution modules when the network goes deeper,}} can have better performance than the U-net structure. Future work will focus on designing and testing different network structures to better improve the results {\txtc{as well as testing the robustness of the trained network to diseased data sets.}}
\section*{References}

\end{document}